\documentclass[aps,pre,twocolumn,superscriptaddress,nolongbibliography]{revtex4-2}  % for review and submission
\usepackage{graphicx}  % needed for figures
\usepackage{dcolumn}   % needed for some tables
\usepackage{bm}        % for math
\usepackage{amssymb}   % for math
\usepackage{hyperref}
\usepackage{physics}
\usepackage{algorithm,algpseudocode}
\usepackage{xcolor}

\begin{document}

% the following line is for submission, including submission to the arXiv!!
%\hspace{5.2in} \mbox{Fermilab-Pub-04/xxx-E}

\title{Interplay between percolation and glassiness in the random Lorentz gas}

% authors
\author{Giulio Biroli}
\affiliation{Laboratoire de Physique de l'Ecole Normale Sup\'erieure, ENS, Universit\'e PSL, CNRS, Sorbonne Universit\'e, Universit\'e de Paris, F-75005 Paris, France
}
\author{Patrick Charbonneau}
\affiliation{Department of Chemistry, Duke University, Durham, North Carolina 27708, USA}
\affiliation{Department of Physics, Duke University, Durham, North Carolina 27708, USA}
\author{Eric I.~Corwin}
\affiliation{Department of Physics and Material Science Institute, University of Oregon, Eugene, Oregon 97403, USA}
\author{Yi Hu}
\email{yi.hu@duke.edu}
\affiliation{Department of Chemistry, Duke University, Durham, North Carolina 27708, USA}
\author{Harukuni Ikeda}
\affiliation{Graduate School of Arts and Sciences, The University of Tokyo 153-8902, Japan
}
\author{Grzegorz Szamel}
\affiliation{Department of Chemistry, Colorado State University, Fort Collins, CO 80523, USA}
\author{Francesco Zamponi}
\affiliation{Laboratoire de Physique de l'Ecole Normale Sup\'erieure, ENS, Universit\'e PSL, CNRS, Sorbonne Universit\'e, Universit\'e de Paris, F-75005 Paris, France
}

\date{\today}

\begin{abstract}
The random Lorentz gas (RLG) is a minimal model of transport in heterogeneous media. It also models the dynamics of a tracer in a glassy system.
These two perspectives, however, are fundamentally inconsistent. Arrest in the former is related to percolation, and hence continuous, while glass-like arrest is discontinuous. 
In order to clarify the interplay between percolation and glassiness in the RLG, we consider its exact solution in the infinite-dimensional $d\rightarrow\infty$ limit, as well as numerics in $d=2\ldots 20$. 
We find that the mean field solutions of the RLG and glasses fall in the same universality class, and that instantonic corrections related to rare cage escapes destroy the glass transition in finite dimensions. This advance suggests that the RLG can be used as a toy model to develop a first-principle description of hopping in structural glasses.
\end{abstract}

\maketitle

\section{Introduction}

The random version of the venerable Lorentz gas (RLG) consists of a tracer navigating between a collection of Poisson-distributed hard spherical obstacles. Despite the apparent simplicity of the model, its phenomenology is quite rich. As the obstacle density increases, tracer diffusion is first delayed and then suppressed altogether. In physical dimensions, $d=2,3$, the localization transition provably coincides with that of void space percolating~\cite{kertesz1981percolation,elam1984critical}, and is hence continuous and accompanied by an extended subdiffusive regime~\cite{stauffer1994percolation,ben2000diffusion,hofling2006localization}. The minimal complex nature of the RLG makes it a standard model of transport in heterogeneous media for systems as diverse as electrons in metals with impurities~\cite{dmitriev2002anomalous} and proteins in cells~\cite{hofling2013anomalous,treado2019void}.

\begin{figure*}[t]
\includegraphics[width=0.8\textwidth]{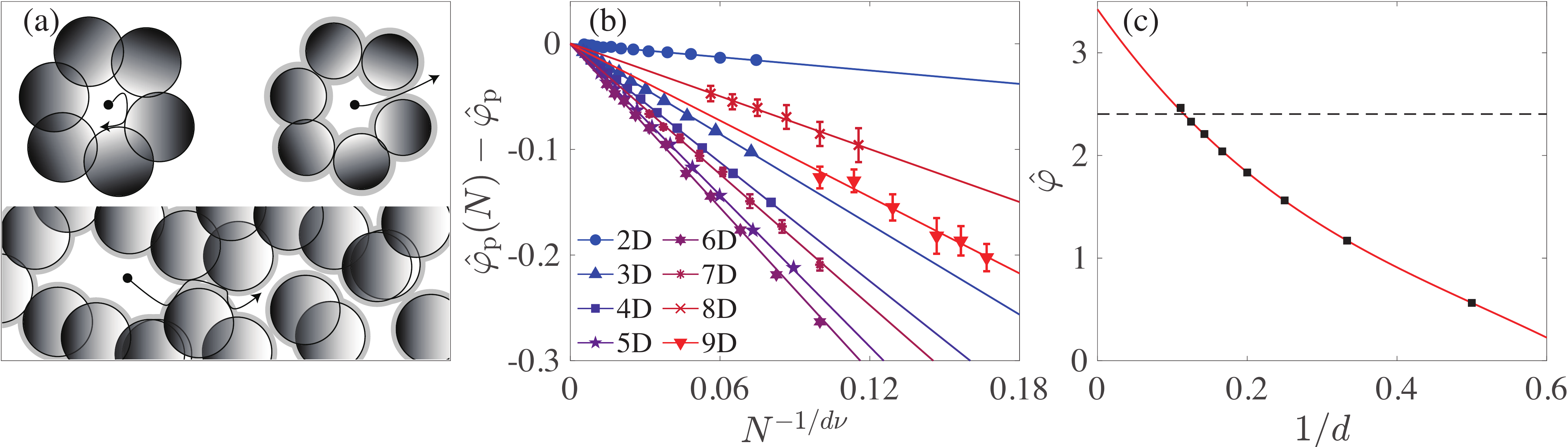}
\caption{(a) Two descriptions of RLG uncaging upon decreasing $\hat\varphi$: (top) the MFT cage discontinuously disintegrates at $\hat\varphi_\mathrm{d}$; (down) cages merge continuously and form an infinite percolating cluster at $\hat\varphi_\mathrm{p}$. (b) Finite-size scaling of $\hat\varphi_\mathrm{p}(N)$ in $d=2$ to $9$.  
(c) $\hat\varphi_\mathrm{p}(\infty)$ in $d=2$ to $9$ (squares, from right to left) compared with the MFT $\hat\varphi_\mathrm{d}$ (dashed line). The red line denotes a polynomial fit to the percolation thresholds, Eq.~\eqref{eq:fit}. Extrapolating this form strongly suggests that $\hat\varphi_\mathrm{p}(d \rightarrow \infty) \neq \hat\varphi_\mathrm{d}$.}
\label{fig:thresholdscaling}
\end{figure*}

The RLG also plays a key role in the theory of glasses. Its consideration was an important step toward formulating the mode-coupling theory (MCT) of glasses~\cite{gotze1981dynamical,leutheusser1984dynamical,szamel2004gaussian,jin2015dimensional}, and it has provided key insight into the role of pinning particles in deeply supercooled liquids~\cite{krakoviack2007mode,kim2009slow,kurzidim2009single,szamel2013glassy}.  The RLG is additionally related to a special limit of the non-convex perceptron, which is a minimal model for glasses and jamming~\cite{franz2016simplest}. 
Further insight into the model arises from noting that the RLG can be construed as a special limit of a hard sphere binary mixture~\cite{coluzzi1999thermodynamics,biazzo2009theory,ikeda2016note}, with one component --the obstacles-- being infinitely smaller than the infinitely-dilute other --the tracer. (Exchanging obstacle and tracer sizes recovers Fig.~\ref{fig:thresholdscaling}(a)~\cite{jin2015dimensional}.) The RLG model should thus be part of the hard sphere glass universality class, and thus similarly undergo a discontinuous dynamical caging transition~\cite{francesco2020theory}.
Interestingly, one could argue that, while in finite-dimensional glass formers this dynamical transition is avoided because various intervening {\it collective} activated processes, including nucleation~\cite{dzero2005activated} and facilitation~\cite{berthier2011theoretical}, the RLG, which by construction eliminates all such effects, should exhibit this transition more crisply.

A paradox, however, follows from this reasoning (Fig.~\ref{fig:thresholdscaling}(a)). On the one hand, the exact mapping of the RLG to a percolation transition gives rise to a \emph{continuous} localization transition~\cite{stauffer1994percolation,ben2000diffusion,hofling2006localization,jin2015dimensional}; on the other hand, the analogy to glass formation gives rise to a \emph{discontinuous} caging transition, at least in the high dimensional, $d\rightarrow\infty$ limit where such description is exact, as we show below. The simplest possible resolution of this paradox, namely that the nature of the percolation transition might change in the $d \rightarrow \infty$ limit, was recently ruled out~\cite{biroli2019dynamics}. Could it then be that the large asymmetry limit of binary hard spheres is singular? Or that the $d\rightarrow\infty$ limit is pathological in some unexpected way?

In this Rapid Communication, we answer these questions and resolve the interplay between percolation and glassiness in the RLG. In Section~\ref{sec:mft}, we first formulate a mean-field theory (MFT) description of the RLG independent of the binary mixture analogy to validate its premise, and then numerically obtain accurate percolation thresholds in Sec.~\ref{sec:phip}, as well as static and dynamical descriptions of caging in Sec.~\ref{sec:csize} and~\ref{sec:dyn}, respectively. Most importantly, by considering spatial dimensions high enough for glassiness to markedly impact the RLG dynamics, we uncover activated microscopic processes that are expected to play a key role in glass formation and yet have thus far eluded theoretical grasp. We thus conclude in Sec.~\ref{sec:conclusion} that the RLG captures some of the key activated processes in glasses, and is simple enough to be treated analytically and numerically, thus opening the way for a first-principle description of such processes.

\section{Mean-field theory derivation} \label{sec:mft}

The MFT of glass-forming liquids, which becomes exact in the $d\to\infty$ limit~\cite{francesco2020theory}, predicts the existence of a dynamical (MCT-like) transition, at which the long-time limit of the scaled mean squared displacement (MSD), $\hat\Delta = d \Delta$, jumps from diverging diffusively to a finite value. Because the infinitely-asymmetric binary fluid mixture that coincides with the RLG in the $d\rightarrow\infty$ limit might however be singular, we here sidestep this reasoning by solving the model directly by cavity reconstruction.
Writing the explicit partition function for the RLG and using the replica symmetric construction of Refs.~\onlinecite{parisi2010mean,francesco2020theory,SI}, one obtains a self-consistent expression for $\hat\Delta$
\begin{equation} \begin{aligned}
\frac{1}{2 \hat\varphi} = -\hat\Delta \int_{-\infty}^{\infty} \dd h e^h \log q(\hat\Delta/2, h) \pdv{q(\hat\Delta/2, h)}{\hat\Delta},
\end{aligned}\label{eq:delta} \end{equation}
at the dimensionally-rescaled packing fraction $\hat\varphi=\rho V_d/d$, where $\rho$ is the number density of obstacles,  $V_d$ is the volume of $d$-dimensional unit sphere~\cite{SI}, and $q(\hat\Delta, h) = \{1 + \erf[ (h + \hat\Delta/2)/\sqrt{2 \hat\Delta} ]\}/2$.
Equation~\eqref{eq:delta} gives that a dynamical glass transition takes place at  $\hat\varphi_\mathrm{d}= 2.4034\ldots$, half that for $d\rightarrow\infty$ hard spheres~\cite{parisi2010mean}.

Considering that $\hat{\Delta}$ is an order parameter for both the percolation and the glass transitions, one may expect the theory of glasses to also describe percolation criticality. This is not the case. While the cage size is expected to diverge logarithmically in mean-field percolation~\cite{biroli2019dynamics},
we here obtain that the cage size is twice that of hard spheres, i.e., $\hat\Delta=2 \hat{\Delta}_\mathrm{HS}(2 \hat\varphi)$, and thus also presents a square-root singularity upon approaching $\hat\varphi_\mathrm{d}$, i.e., $\hat{\Delta}(\hat\varphi_\mathrm{d}) - \hat{\Delta}(\hat\varphi) \sim \sqrt{\hat\varphi - \hat\varphi_\mathrm{d}}$. 
In other words, the RLG and hard spheres share a same MFT universality class characterized by a discontinuous glass transition, that is distinct from percolation.

\section{Percolation threshold} \label{sec:phip}

Although the percolation criticality is distinct from that of the dynamical glass transition, one might nonetheless wonder whether the former smoothly extrapolates to latter in the limit $d\to\infty$. We thus consider the scaling of the percolation threshold, $\hat\varphi_\mathrm{p}$, with dimension to determine if it coincides with the MFT prediction for $\hat\varphi_\mathrm{d}$ in the $d\rightarrow\infty$ limit. In systems with $N$ Poisson-distributed obstacles in a $d$-dimensional box under  periodic boundary conditions, the void percolation can be mapped onto the bond percolation of a network built on the Voronoi tessellation of obstacles~\cite{kerstein1983equivalence}, assigning to each edge of that tessellation the smallest obstacle radius $\sigma$ that can block it, and using a disjoint-set forest algorithm adapted from continuum-space analysis to identify the percolated cluster~\cite{newman2001fast,mertens2012continuum}. Optimizing the periodic boundary conditions~\cite{convay1982fast,SI} and the Voronoi tessellation~\cite{boissonnat2005convex,Quickhull2016,SI} enables us to obtain $\hat\varphi_\mathrm{p}(N)$ up to $d=9$. The thermodynamic $\hat\varphi_\mathrm{p}$ is then extracted by fitting (Fig.~\ref{fig:thresholdscaling}(b))
\begin{equation}
|\hat\varphi_\mathrm{p}(N) - \hat\varphi_\mathrm{p}| \sim N^{-1/d \nu},
\end{equation}
where $\nu$ is the percolation correlation length exponent~\cite{koza2016discrete,SI}. 

For $d\leq 8$, $\hat\varphi_\mathrm{p}<\hat\varphi_\mathrm{d}$, but $\hat\varphi_\mathrm{p}(d=9) = 2.46(4) > \hat\varphi_\mathrm{d}$, indicating that the order of the two switches between $d=8$ and $9$. Fitting the results to a cubic form,
\begin{equation}
\label{eq:fit}
\hat\varphi_\mathrm{p} = 3.42(8) - 10.3(9)\frac{1}{d} + 13(3)\left(\frac{1}{d}\right)^2 -9(4)\left(\frac{1}{d}\right)^3,
\end{equation}
further gives $\hat\varphi_\mathrm{p}(d \rightarrow \infty)=3.42(8)$, which differs significantly from the MFT prediction (Fig.~\ref{fig:thresholdscaling}(c)).
In other words, while MFT is expected to be exact in $d \rightarrow \infty$ limit, it fails to capture the percolation transition in that same limit, which only heightens the paradox.
For $d<8$ the tracer is localized on both sides of $\hat\varphi_\mathrm{d}$, and hence the dynamical transition has no physical meaning, but a densifying system in $d>8$ might first encounter around $\hat\varphi_\mathrm{d}$ (imperfect) local cages that collectively percolate and can be escaped via activated processes~\cite{charbonneau2014hopping}, before being properly localized at the percolation threshold $\hat\varphi_\mathrm{p}$. 

\section{Cage Sizes} \label{sec:csize}

In order to ascertain this scenario, the MFT description of caging  needs first to be assessed. To do so, we implement a cavity reconstruction scheme adapted from Refs.~\cite{sastry1997statistical,charbonneau2014hopping,SI}, which can be viewed as the continuum-space generalization of the Leath algorithm~\cite{leath1976clustersize}. 
Specifically, we define a hyperspherical shell, centered at the origin, of inner radius $\sigma$ and outer radius $r_\mathrm{max}$, and pick a number of obstacles $N$ from the Poisson distribution $p(N) = N_0^N e^{-N_0}/N!$ with $N_0 = d \hat\varphi (r^d_\mathrm{max} - \sigma^d)$, which are then placed uniformly at random within that shell. (The choice of $r_\mathrm{max}$ is such that the cavity containing the origin is closed.)
This algorithm guarantees that the probability of obtaining a cavity containing the origin, $\mathbb{C}$, exactly tracks the distribution of cavities at that same $\hat\varphi$ in an infinitely large system. A set of randomly distributed points $\{S_i\}$, within $\mathbb{C}$ can then be used to compute the second moment of the coordinates,
\begin{equation}
\Delta(\mathbb{C}) = \langle (S_i - S_j)^2 \rangle = 2 (\langle S_i^2 \rangle - \langle S_i \rangle^2 ),
\end{equation}
and then $\Delta=\mathbb{E}_\mathbb{C}[\Delta(\mathbb{C})]$. Physically, this method provides the long-time limit of the MSD of a tracer without explicitly running its dynamics, which is advantageous because it eliminates putative dynamical bottlenecks. However, because its computational cost increases exponentially with $d$, for $d \ge 8$ the explicit long-time limit of the tracer dynamics needs to be computed to estimate $\Delta$. The agreement between the two approaches at intermediate $d$ indicates that bottlenecks can be confidently neglected in this regime.

\begin{figure}[t]
\includegraphics[width=1\columnwidth]{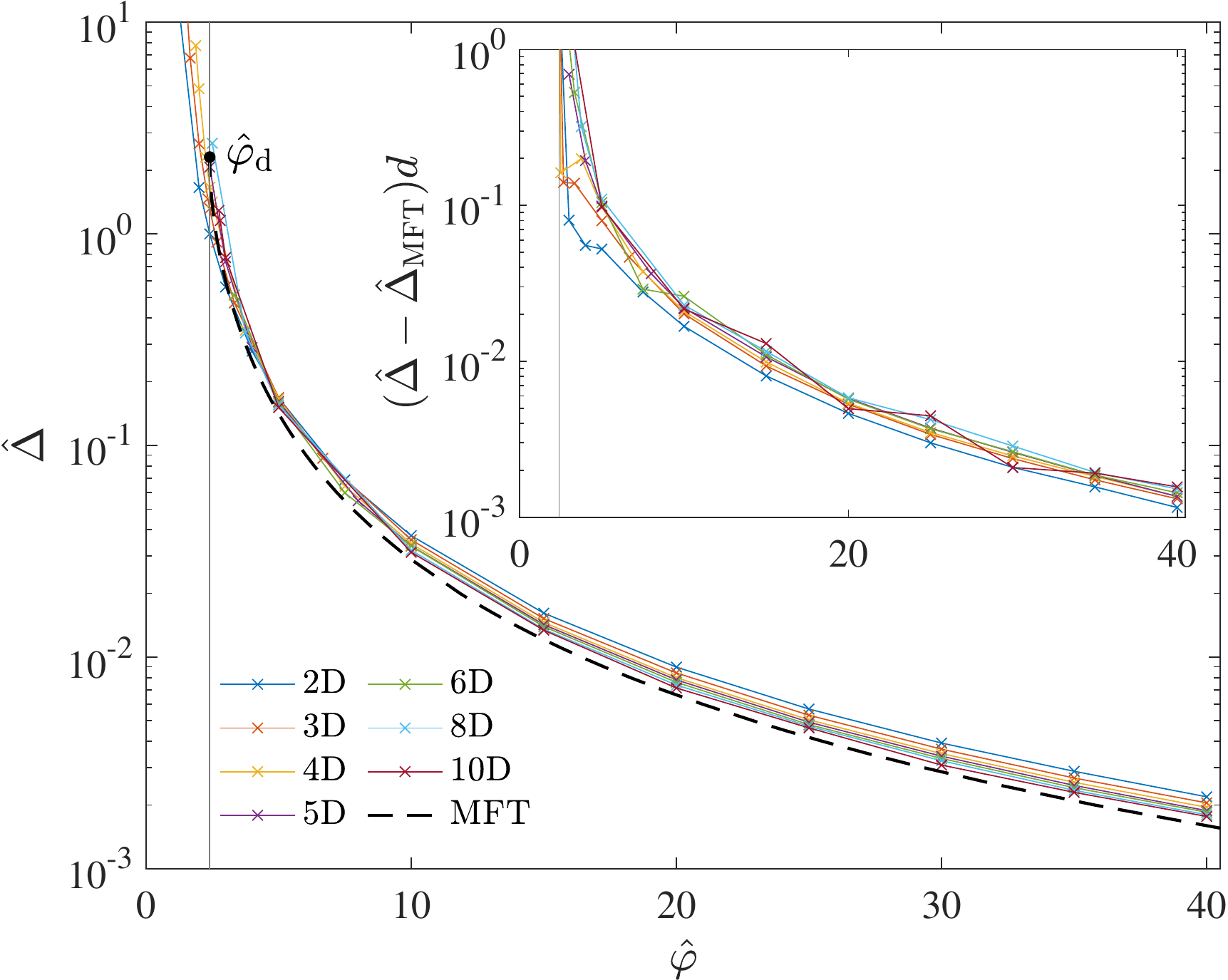}
\caption{Scaling of the cage size with density for different $d$. Results for $d=2$ to $6$ are obtained by random sampling, and those for $d\geq8$ from the long-time caging dynamics. At high densities, the MFT, $d\rightarrow\infty$ prediction (dashed line) is steadily approached as $d$ increases, but in small densities the percolation criticality dominates the growth of cage size. Note that results in high dimension and small densities are numerically inaccessible. (inset) The scaling collapse of the deviation from the MFT prediction identifies the dominant $1/d$ correction.}
\label{fig:cagesize}
\end{figure}

For $\hat\varphi\gg\hat\varphi_\mathrm{d}$, the (scaled) cage size nicely converges to the MFT prediction as $d$ increases (Fig.~\ref{fig:cagesize}), and the dominant correction is perturbative in $1/d$.  In this high-density regime, the quantitative accordance with  MFT is robust down to physical dimensions. A generalized MFT with perturbative corrections should thus offer accurate predictions in all $d$, a clear opportunity for future theoretical studies.

By contrast, for $\hat\varphi\sim\hat\varphi_\mathrm{d}$, a regime dominated by percolation criticality---with $\Delta$ diverging at $\hat\varphi_\mathrm{p}$---is observed (Fig.~\ref{fig:cagesize}). The static cage size either crosses $\hat\varphi_\mathrm{d}$ smoothly or is expected to diverge before reaching $\hat\varphi_\mathrm{d}$ from above, depending on the relative order of $\hat\varphi_\mathrm{d}$ and $\hat\varphi_\mathrm{p}$. These strong discrepancies with respect to MFT found around $\hat\varphi_\mathrm{d}$ hint at a complex interplay between glass and percolation physics. 

\section{Tracer Dynamics} \label{sec:dyn}

\begin{figure}[t]
\includegraphics[width=1\columnwidth]{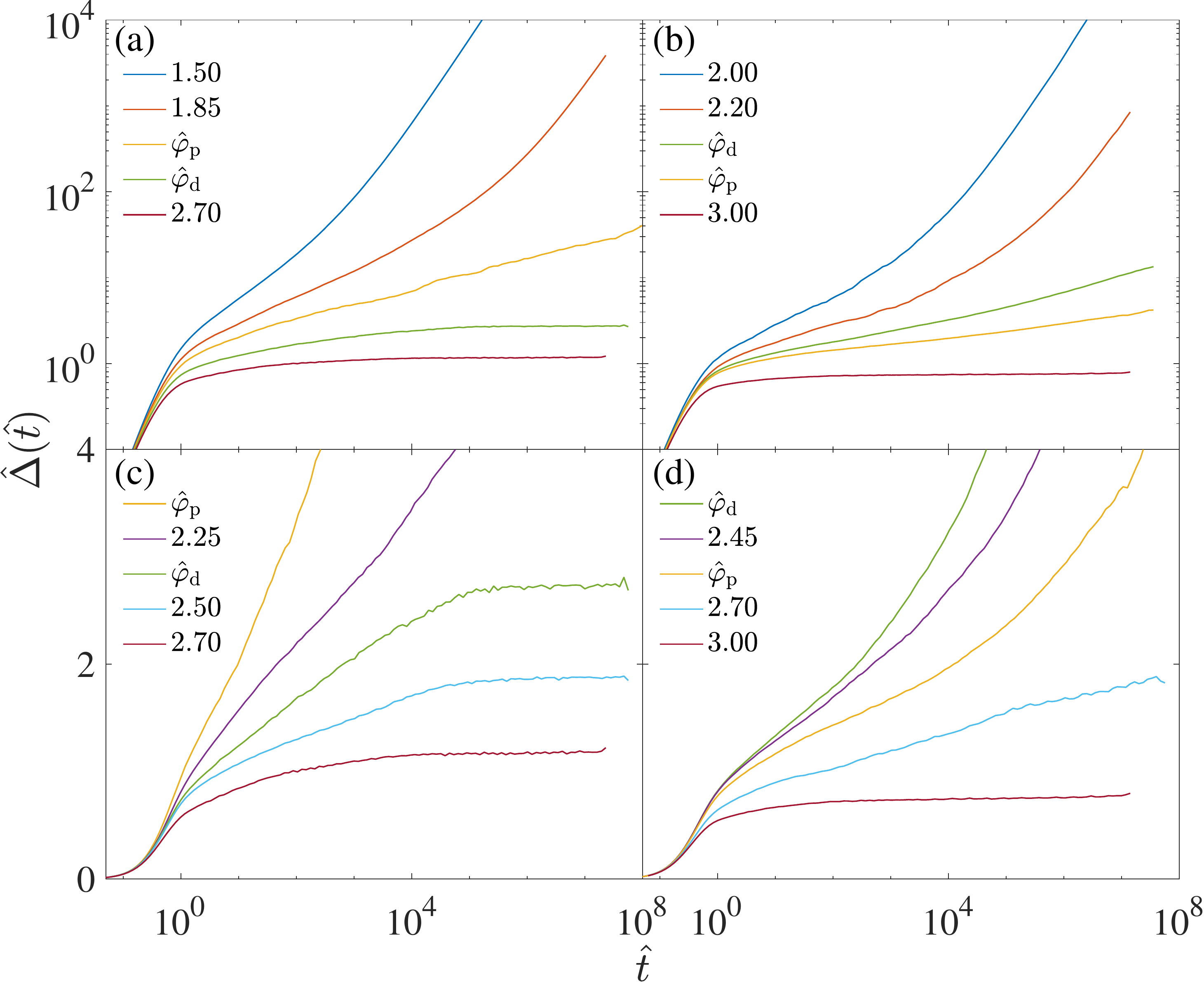}
\caption{Time evolution of the MSD in the ballistic dynamics in (a-b) $d=6$ and $10$ under log-log scale, and (c-d) same dimensions under log-lin scale. The long time dynamics is diffusive for $\hat\varphi < \hat\varphi_\mathrm{p}$ or localized for $\hat\varphi > \hat\varphi_\mathrm{p}$. In $d \ge 6$, $\hat\Delta$ at $\hat\varphi_\mathrm{p}$ is expected to grow logarithmically. In (d), specifically, the signature of intermediate dynamical slowdown emerges before the logarithmic growth.
}
\label{fig:dynamics}
\end{figure}

In order to disentangle the two, we consider the dynamical counterpart of the above static description. We first examine the tracer dynamics, following the ballistic approach of H{\"o}fling \emph{et al.}~\cite{hofling2006localization,hofling2008critical}, but setting the microscopic timescale such that the short-time growth of the MSD scales as $\hat\Delta(t) = \hat{t}^2$ when $\hat{t} \rightarrow 0$ in all dimensions.
As expected from percolation theory~\cite{ben2000diffusion,biroli2019dynamics}, in the long-time limit either localization or diffusion is observed,  for $\hat\varphi > \hat\varphi_\mathrm{p}$ and $\hat\varphi < \hat\varphi_\mathrm{p}$, respectively (Fig.~\ref{fig:dynamics}(a,b)). An intermediate subdiffusive regime, which scales logarithmically with time for $d\geq6$~\cite{biroli2019dynamics}, also develops around the percolation threshold, and fully dominates the dynamics at $\hat\varphi=\hat\varphi_\mathrm{p}$.
Figure~\ref{fig:dynamics}(c,d) considers more closely the interplay between $\hat\varphi_\mathrm{p}<\hat\varphi_\mathrm{d}$. In $d=6$, no hint of MFT-like caging is observed around $\hat\varphi_\mathrm{p}$, as expected. Because $\hat\varphi_\mathrm{p}<\hat\varphi_\mathrm{d}$, percolation dominates the caging dynamics. Hence, for $\hat\varphi > \hat\varphi_\mathrm{p}$, logarithmic growth immediately follows the ballistic regime until a plateau is reached.  By contrast, in $d=10$, where $\hat\varphi_\mathrm{p}>\hat\varphi_\mathrm{d}$, a weak dynamical slowdown emerges at intermediate times for $\hat\varphi \ge \hat\varphi_\mathrm{d}$. Such a pre-asymptotic effect is distinctly absent in lattice systems~\cite{biroli2019dynamics}. However, conclusively determining whether this slowdown is controlled by MFT caging or by some other model-specific pre-asymptotic correction to percolation criticality would require higher-dimensional systems than this computational scheme currently permits.

\begin{figure}[t]
\includegraphics[width=1\columnwidth]{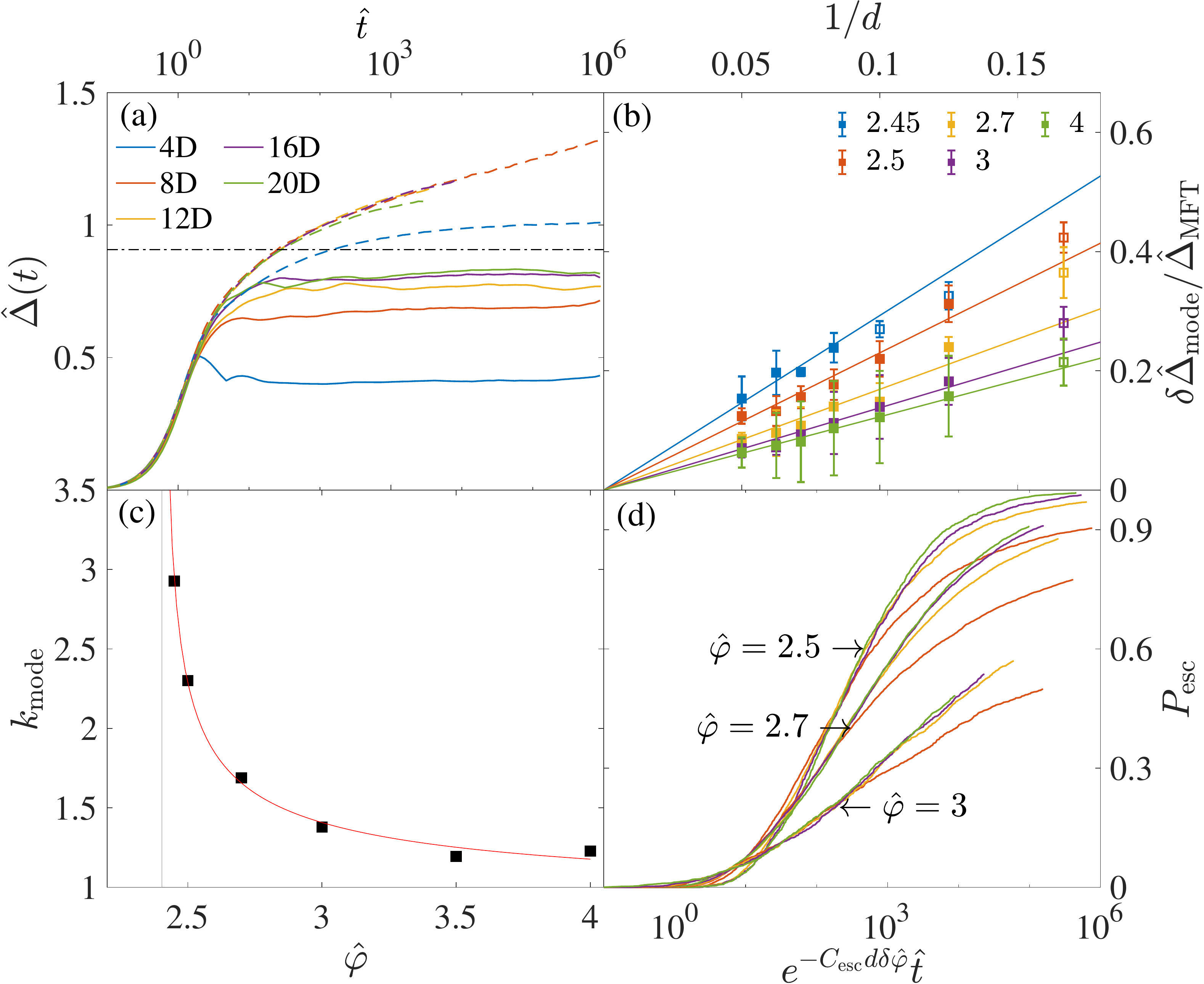}
\caption{Cages and cage escapes in $d=4$ to $20$ obtained from dynamical cavity reconstructions.  (a) Modal (solid line) and mean (dashed line) squared displacements of tracers with time at $\hat\varphi = 2.7$, along with the MFT prediction (dash-dotted line). MSD curves terminate when $2\%$ of tracers have escaped. While the MSD drifts with time, the mode robustly plateaus. (b) The plateau of $\hat\Delta_\mathrm{mode}$ approaches the MFT prediction as in Eq.~\eqref{eq:modcorr}  
for various $\hat\varphi$; (c) the scale of the perturbative correction to the cage size grows upon approaching $\hat{\varphi}_\mathrm{d}$, and empirically fits $k_\mathrm{mode} = 0.46/\sqrt{\delta\hat\varphi}+0.81$. (d) Cage escape probabilities for $\hat\Delta_\mathrm{esc} = 4$ in $\hat\varphi = 2.5, 2.7$ and $3$ collapse under an instantonic form with empirical prefactor $C_\mathrm{esc}=0.4$.}
\label{fig:dytypical}
\end{figure}

To pinpoint the origin of this weak dynamical slowdown, we instead seek an observable more sensitive to MFT-like caging. Recalling that percolation criticality is dominated by rare large cages, while MFT is evaluated via a saddle point that extracts the \emph{typical} cage size, we choose to focus on the \emph{modal} cage size, i.e.,  $\hat\Delta_\mathrm{mode}(t) = \arg\max P(\hat\Delta(t))$~\cite{SI}. 
By construction, $\hat\Delta_\mathrm{mode}$ eliminates the contribution of rare large cages and cage escapes, and thus effectively plays the same role as the generalized MSD considered in recent glass studies~\cite{charbonneau2014hopping,berthier2019finite}.
This observable is further amenable to a dynamical version of the static cavity reconstruction. Although this setup misses finite-yet-large cages, it provides a sufficiently broad span of the cage-size distribution to reliably identify $\hat\Delta_\mathrm{mode}$. It also extends the numerically accessible dimensional range.
Results up to $d = 20$ and averaged over at least $2 \times 10^3$ independent samples with $\hat\Delta_\mathrm{max} = d \cdot (r_\mathrm{max} - \sigma)^2 \ge 14$ are reported in Fig.~\ref{fig:dytypical}. We find that $\hat\Delta_\mathrm{mode}$ plateaus quickly after the ballistic regime, even near $\hat\varphi_\mathrm{p}$, and that this plateau steadily approaches the MFT caging prediction as $d$ increases (Fig.~\ref{fig:dytypical}(a)). Note that a small finite-size corrections due to finite shell thickness appear in $d=20$, but remain within the statistical error range~\cite{SI}.
Remarkably, the approach to the MFT prediction exhibits a perturbative $1/d$ correction (Fig.~\ref{fig:dytypical}(b)), 
\begin{equation}
\hat\Delta = \hat\Delta_\mathrm{MFT} - \frac{k_\mathrm{mode}}{d}, 
\label{eq:modcorr}
\end{equation}
even fairly close to $\hat\varphi_\mathrm{d}$.  Because the correction prefactor, $k_\mathrm{mode}$,  increases as $\delta\hat\varphi = \hat\varphi-\hat\varphi_\mathrm{d}$ shrinks, we further obtain that perturbative corrections to MFT become increasingly pronounced upon approaching $\hat\varphi_\mathrm{d}$.  Two processes beyond the $d\rightarrow\infty$ MFT description, however, also then appear: (i) the cage size distribution displays a large-$\hat\Delta$-tail, and (ii) a substantial fraction of tracers escape the shell.  As a result, within the range of system sizes and dimensions accessible in numerical simulations, the mode no longer converges for $\hat\varphi < 2.45$.

In order to disentangle these two different physical contributions and to resolve how the MFT description emerges as $d$ increases, we consider the first-passage time of the tracer escaping from a center square distance, $\hat{\Delta}_\mathrm{esc}$. For a fixed scaled density $\hat\varphi > \hat\varphi_\mathrm{d}$, the onset of cage escapes is found to be exponentially delayed in time with increasing dimension for $d \ge 8$ (Fig.~\ref{fig:dytypical}(d)). More precisely, the cumulative probability of a tracer escaping, $P_\mathrm{esc}(\hat{t})$, at fixed $\hat\varphi$ follows a scaling form 
\begin{equation} \label{eq:instantonic}
P_\mathrm{esc}(\hat{t}; \delta\hat\varphi) \sim \hat{f}(e^{-C_\mathrm{esc} d \delta\hat\varphi} \hat{t}; \delta \hat\varphi),
\end{equation}
with master function $\hat{f}(x;\delta\hat\varphi)$ and a prefactor  $C_\mathrm{esc}(\hat\Delta_\mathrm{esc})\approx 0.4$ that depends only weakly on the choice of cutoff for $\hat\Delta_\mathrm{esc}/\hat\Delta_\mathrm{mode}\sim\mathcal{O}(1)$. In small dimensions, however, cage escapes deviate from this scaling form. Mean-field--like caging around $\hat\varphi_\mathrm{d}$ is then so weak that higher-order corrections dominate.

We can now properly understand the logarithmic drift of the MSD that appears at intermediate times when  $\hat\varphi_\mathrm{p}>\hat\varphi_\mathrm{d}$ as being due to imperfect caging. As dimension increases, the MFT caging prediction is recovered because the prefactor of the logarithm slowly vanishes. Geometrically, most cages are open for $\hat\varphi_\mathrm{d} <\hat\varphi < \hat\varphi_\mathrm{p}$, thus giving rise to void percolation, but escape paths out of open cages steadily shrink with increasing $d$, giving rise to more pronounced dynamical caging. This collapse form further suggests that near $\hat\varphi_\mathrm{d}$ cage escapes are so prevalent that they dominate the dynamics in any finite $d$.  Such {\it hopping} processes (exponentially suppressed in $d$ by contrast to $1/d$ perturbations) have long been debated in glass physics~\cite{gotze1988scaling,bhattacharyya2005bridging,mirigian2014elastically}, but this particular instantonic correction to the MFT of glasses was previously unknown. More than a mere correction, it is here found to be the primary reason why the sharp mean-field dynamical glass transition becomes a crossover in finite $d$.

\section{Conclusion} \label{sec:conclusion}

We have analyzed the interplay between glassiness and percolation in the RLG, and obtained quantitative evidence of non-trivial finite-dimensional corrections to MFT. More specifically, we have found that the static cage size at high density and the typical dynamical cage size at all densities show a \emph{perturbative}, $1/d$, correction to the MFT $d \rightarrow \infty$ result, and that \emph{non-perturbative} dynamical cage escapes are suppressed exponentially with $d$ around $\hat\varphi_\mathrm{d}$. In the RLG, these finite-dimensional corrections are dominant in physical dimensions, $d=2,3$. Our work therefore reveals in a precise and concrete way the important role played by activated processes in avoiding the dynamical glass transition. 

Having identified these two types of corrections to MFT that go beyond the traditional instantonic picture~\cite{dzero2005activated} and facilitation~\cite{berthier2011theoretical}, we should now be able to identify activated processes for more realistic models of glasses and obtain first-principle description of non-perturbative corrections to MFT for finite-dimensional disordered systems. Our results also offer a putative first-principle pathway for relating local structure and dynamics in glass-forming liquids~\cite{royall2015role}.

\begin{acknowledgments}
We thank Antonio Auffinger, Benoit Charbonneau, Sayan Mukherjee, Giorgio Parisi, and Alexander Reznikov for stimulating discussions. This work was supported by a grant from the Simons Foundation (\#454937, Patrick Charbonneau; \#454939, Eric Corwin; \#454935, Giulio Biroli; \#454955, Francesco Zamponi). This research was also supported in part by the National Science Foundation under Grant No. NSF PHY-1748958.  The computations were carried out on the Duke Compute Cluster and Open Science Grid~\cite{osg07,osg09}, supported by National Science Foundation award 1148698, and the U.S. Department of Energy's Office of Science. Data relevant to this work have been archived and can be accessed at the Duke Digital Repository.
\end{acknowledgments}

\appendix

\section{Notation} \label{sec:notation}
In order to investigate the interplay beetween the percolation and glassiness in the RLG, we first need to reconcile the two set of notations. The central quantity for both is the number density of obstacles, $\rho$, which allows to define a unitless volume fraction of obstacles $\Phi = \rho V_d \sigma^d$, where $V_d$ is the $d$-dimensional volume of a unit sphere and $\sigma$ is the obstacle radius. For the RLG, the obstacle radius is commonly set to $\sigma=1$ while the tracer radius $\sigma_\mathrm{tracer}$ is infinitesimal, and hence naturally we can define $\Phi = \rho V_d$.
Without loss of generality, and by analogy to the Mari-Kurchan model~\cite{mari2009jamming,charbonneau2014hopping,jin2015dimensional}, we can equivalently choose $\sigma=\sigma_\mathrm{tracer}=1/2$, which naturally defines $\varphi = \Phi 2^{-d}$.
For high-dimensional scaling convenience, we further define the rescaled packing fraction
\begin{equation}
\hat\varphi = \Phi/d = 2^d \varphi /d.
\end{equation}
Similarly, the cage size, $\Delta$, defined as the infinite-time limit of mean squared displacement (MSD) of the tracer, can be rescaled as $\hat\Delta = d \cdot \Delta$.
For reference, Table~\ref{tab:notation} provides the correspondence between notations commonly used in the scientific literature about the RLG.

\begin{table}[ht]
\caption{Common notations for packing fraction and cage size}
\begin{tabular}{ccc}
\hline \hline
Quantity & Equivalence \\
\hline
$\rho$~\cite{jin2015dimensional} & $n$~\cite{hofling2006localization} \\
$\Phi$~\cite{jin2015dimensional} & \\
$\varphi$~\cite{jin2015dimensional,francesco2020theory} & \\
$\hat\varphi$~\cite{charbonneau2014hopping,jin2015dimensional,francesco2020theory} & \\
$\Delta$~\cite{charbonneau2014hopping} & $\delta r^2/\sigma^2$~\cite{hofling2006localization} \\ 
$\hat\Delta$ & $\Delta$~\cite{francesco2020theory}, $Ad^2$~\cite{mangeat2016quantitative} \\
\hline
\end{tabular}
 \label{tab:notation}
\end{table}

\section{Mean-field Theory Derivation}
As mentioned in the main text, the RLG cage size in the $d\rightarrow\infty$ limit can be obtained by applying the replica technique to a cavity computation. We here provide details about this derivation. The setup consists of $N$ hard spherical obstacles of radius $\sigma$ placed uniformly at random at positions $R_i$ within a volume $V$ centered around the origin. The free volume available to a tracer placed at the origin is thus
\begin{equation}
Z = \int \dd x \prod_{i=1}^N \theta( |x - R_i| > \sigma) \ ,
\end{equation}
where $\theta(x)$ denotes the Heaviside function. 
The replicated partition function in the thermodynamic, $N \to \infty$, limit at fixed obstacle density $\rho = N/V$ is  
\begin{equation}
\overline{Z^n} = \int \dd \overline x  \left[ \frac{ \int_{|R|>\sigma} \dd R \prod_{a=1}^n \theta( |x_a - R| > \sigma)}{ \int_{|R|>\sigma} \dd R} \right]^N,
\end{equation}
and the free energy is
\begin{equation}
F = - \overline{\log Z} = - \lim_{n\to 0} \partial_n \overline{Z^n}.
\end{equation}

For this system, we expect two phases:
\begin{itemize}
\item In the \emph{liquid} phase, the (replicated) tracers are not confined close to the origin. Each replica thus decorrelates over the whole volume and 
\begin{equation} \begin{aligned}
\overline{Z^n} &\sim V^n \left(\frac{V - (n+1) V_\sigma}{V-V_\sigma}\right)^N \sim V^n e^{-n \rho V_\sigma}, \\
F_{\rm liq} &= -\log V + \rho V_\sigma \ ,
\end{aligned} \end{equation}
where $V_\sigma$ is the volume of a sphere of diameter $\sigma$.
\item In the \emph{glass} phase, with high probability the origin is contained within a cage. Many other cages exist in the volume, but a tracer starting at the origin remains confined within that cage. Note that the cage at the origin is metastable, because faraway cages thermodynamically dominate the measure, hence the choice of a cavity computation.
\end{itemize}

In the glass phase, we can write, after introducing a fictitious coordinate $x_0=0$,
\begin{equation} \begin{split}
\overline{Z^n} &=  \int d\overline x  \left\{  \frac{\int \dd R \left[\prod_{a=0}^n \theta( |x_a - R| > \sigma) \right]}{ V- V_\sigma} \right\}^N \\
&= \int \dd \overline x \left\{ 1 + \frac{\int \dd R \,  \left[\prod_{a=0}^n \theta( |x_a - R| > \sigma)  -1 \right] + V_\sigma }{V-V_\sigma} \right\}^N \\ 
&= \int \dd \overline x \left\{ 1 + \frac{-\int \dd R \,  \theta( \min_{a\in[0,n]} |x_a - R| < \sigma) + V_\sigma}{V-V_\sigma} \right\}^N \\ 
&=  e^{\rho V_\sigma} \int \dd \overline x \ e^{ - \rho \int \dd R  \, \theta( \min_{a\in[0,n]} |x_a - R| < \sigma) } \\ 
& =  e^{\rho V_\sigma} C_{n+1,d} \int \dd \hat{q} \, e^{ \frac{d-n-1}2 \log\det \hat{q} + \rho \overline f_{n+1}(\{0,\overline x\})}, 
\end{split}\end{equation}
where overlap variables have been changed to rotationally invariant quantities, $\hat{q}_{ab}= x_a \cdot x_b$ as in Ref.~\onlinecite[Eq.(2.96)]{francesco2020theory}, and
\begin{equation}
\overline f_n(\overline x) = - \int \dd R \, \theta( \min_{a\in[1,n]} |x_a - R| < \sigma),
\end{equation}
as in Ref.~\onlinecite[Eq.(30)]{kurchan2012exact}.

Following the approach of Refs.~\onlinecite{parisi2010mean,francesco2020theory} for evaluating $\overline{Z^n}$ by saddle point integration for $d \rightarrow \infty$, we obtain
\begin{equation} \begin{aligned}
\log \overline{Z^n} &= \mathrm{cnst} + \frac{d}{2} \log \det \hat{q} + d \hat\varphi \bar{g}_{n} \\
&=  \mathrm{cnst} + \frac{d}{2} \log \left( \left(n \hat\Delta_\mathrm{r} - (n-1) \frac{\hat\Delta}{2}\right)\left(\frac{\hat\Delta}{2}\right)^{n-1} \right) \\ &+ d \hat\varphi \left(\overline f_{n+1}(\hat\Delta_\mathrm{r}, \hat\Delta)  + 1 \right),
\end{aligned} \end{equation}
where irrelevant constants have been dropped and the rescaled squared displacement and density, $\hat{\Delta}$ and $\hat\varphi$, respectively, are defined as in Section~\ref{sec:notation}. By taking the replica symmetric solution, $\hat\Delta_\mathrm{r} = \hat\Delta$, the expression can then be reduced to a one-dimensional integral~\cite{francesco2020theory}, such that
\begin{equation}
 f_{n+1}(\hat\Delta) = \int_{-\infty}^{\infty} \dd h e^h (q(\hat\Delta/2, h)^{n+1} - 1),
\end{equation}
where $q(\hat\Delta, h) = (1 + \erf( \frac{h + \hat\Delta/2}{\sqrt{2 \hat\Delta}} ))/2$. Note that in the original hard sphere derivation, in which all particles oscillate, the large variance term has the form $\hat\Delta = (\hat\Delta_\mathrm{tracer} + \hat\Delta_\mathrm{obstacle})/2$. By contrast, obstacles are fixed in the RLG, and hence $\hat\Delta = \hat\Delta_\mathrm{tracer}/2$.

Under the replica symmetric assumption, the free energy is then
\begin{equation} \begin{aligned}
\log \overline{Z^n} = \mathrm{cnst} + \frac{d n}{2} \log \hat\Delta + d \hat\varphi (\int_{-\infty}^{\infty} \dd h e^h \\ \quad (q(\hat\Delta/2, h)^{n+1} - 1)  + 1).
\end{aligned} \end{equation}
Solving for $\pdv{\log \overline{Z^n}}{\hat\Delta} = 0$ provides the cage size that optimizes the free energy,
\begin{equation} \begin{aligned}
\frac{n}{2 \hat\Delta} = -\hat\varphi \int_{-\infty}^{\infty} \dd h e^h (n+1) q(\hat\Delta/2, h)^n \pdv{q(\hat\Delta/2, h)}{\Delta}.
\end{aligned} \end{equation}
Noting that $\int_{-\infty}^{\infty} \dd h e^h \pdv{q}{\hat\Delta} = 0$ and taking the limit $n \rightarrow 0$, the cage size and the obstacle density are then related by
\begin{equation} \begin{aligned}
\frac{1}{2 \hat\varphi} = -\hat\Delta \int_{-\infty}^{\infty} \dd h e^h \log q(\hat\Delta/2, h) \pdv{q(\hat\Delta/2, h)}{\hat\Delta}.
\end{aligned} \end{equation}

\section{Void Percolation Threshold Computation}
This section details the algorithm used for detecting the void percolation threshold of the RLG. 
We first place $N$ obstacles uniformly at random within a $d$-dimensional box under periodic boundary conditions. Conventional cubic boxes, $\mathbb{Z}^d$, are used in $d \le 6$, while the Wigner-Seitz cell of the checkerboard, $D_d$, lattice, the $E_8$ lattice and $\Lambda_9$ lattice (densest lattice packing in $d=8$ and $9$) are used in $4 \le d \le 7$, $8$ and $9$, respectively. 
A Voronoi tessellation of the obstacles allows us to map the void percolation problem onto that of the bond percolation of edges in that tessellation.
Each edge is weighted by the circumscribed radius of the facet in the Delaunay triangulation that is dual to this edge, which defines the minimum radius of the obstacles that can block this edge.
Because the number of Voronoi vertices and edges grows exponentially with dimension, memory use must be carefully handled. First, we build the tessellation point by point~\cite{charbonneau2013geometrical}.
Specifically, for each obstacle, $p_i$, we calculate the convex hull~\cite{Quickhull2016} of the inverse coordinates of the other obstacles, after translating $p_i$ at the origin. The vertices of this convex hull  then correspond to the neighbors of $p_i$ in the Voronoi tessellation~\cite{boissonnat2005convex}. Second, edges with a sufficiently small weight remain blocked in a percolating network and are dropped on-the-fly, while building the tessellation. Orders of magnitude in memory use are therefore gained, thus enabling the analysis of sufficiently large systems, even in the highest dimension considered.

The percolation threshold is then determined by an algorithm akin to that used for the continuum-space percolation of obstacles~\cite{newman2001fast,mertens2012continuum}. This approach is applied on a disjoint-set forest data structure. A disjoint-set consists of a number of nodes, each of which corresponding to a Voronoi vertex. Each node maintains a parent pointer and the displacement vector to its parent, tracing back to a unique root node in the set. Each disjoint-set thus corresponds to a cavity in the system. Voronoi edges are first sorted in descending order, and then the neighboring vertices of each edge are iteratively considered. If the two vertices, $X_1$ and $X_2$, do not yet belong to a same cavity, they are merged; otherwise, percolation is checked by: 
\begin{enumerate}
\item Calculating the displacement vector between $X_1$ and $X_2$ (under minimal image convention) $\bm{r}_0 = X_1 - X_2$;
\item Calculating the displacement vector from $X_1$ and $X_2$ to the root, $\bm{r}_1$ and $\bm{r}_2$, respectively;
\item Comparing if $\bm{r}_1 - \bm{r}_2 \neq \bm{r}_0$.
\end{enumerate}

If the displacements calculated from the two methods differ (necessarily, by integers), then the cavity must span across the periodic boundary and form a cycle. 
Percolation is deemed to take place when there exist such cycles in all dimensions, which reduces sample-to-sample variations compared to other percolation criteria~\cite{jin2015dimensional}. From the standard percolation universality class~\cite{stauffer1994percolation}, we know that the percolation threshold in a finite system of $N$ obstacles converges to the thermodynamic, $N \rightarrow \infty$, limit, with asymptotic scaling
\begin{equation}
\Phi_\mathrm{p}(N) - \Phi_\mathrm{p}(\infty) \sim N^{-1/d \nu},
\end{equation}
where $\nu$ is the correlation length exponent, $\nu=4/3,0.8774,0.6852,0.5723$ for $d=2$ to $5$~\cite{koza2016discrete} and $1/2$ for d $\ge 6$. Our percolation threshold detection algorithm increases the range of accessible system sizes by orders of magnitude, which makes this fitting robust in all dimensions considered in this work.
Formally, the neighbors of an obstacle obeys the Poisson distribution without bias if the system size (inscribed radius of the periodic box) is greater than the maximum neighbor-distance of obstacles. In high dimension, this condition requires increasingly large number of obstacles in a system. Empirically we find the asymptotic scaling is still recovered in smaller systems, within the range of numerical uncertainty, when the next-nearest periodic image of neighboring obstacles are included in the construction of Voronoi tessellation.
The resulting percolation threshold are listed in Table~\ref{tab:threshold}. Note that our results reveal a systematic bias in the numerical treatment of Ref.~\cite{jin2015dimensional} for $d\geq 4$, because it included pre-asymptotic system sizes in the fit. 

\begin{table}[h]
\caption{Numerical estimates of the void percolation threshold}
\begin{tabular}{cccc}
\hline \hline
$d$ & $\Phi_\mathrm{p}$ & $\hat\varphi_\mathrm{p}$  \\
\hline
2 & 1.1276(9) & 0.5638(5) \\
3 & 3.510(2) & 1.1698(8) \\
4 & 6.248(2) & 1.5621(5) \\
5 & 9.170(8) & 1.834(2) \\
6 & 12.24(2) & 2.040(4) \\
7 &  15.46(5) & 2.209(7)\\
8 & 18.64(8) & 2.330(9) \\
9 & 22.1(4) & 2.46(4) \\
\hline
\end{tabular}
 \label{tab:threshold}
\end{table}

\section{Numerical Cavity Reconstruction Scheme}
\label{sec:cavityreconst}
At high density, the RLG model is amenable to cavity reconstruction, which allows to compute efficiently the properties of the localized regime within a spherical shell of radius $r_\mathrm{max}$. 
The number of obstacles $N$ to be placed within that shell is first picked at random from the Poisson distribution
\begin{equation} \label{eq:poisson}
p(N) = \frac{N_0^N e^{-N_0}}{N!},
\end{equation}
where $N_0 = d \hat\varphi (r^d_\mathrm{max} - \sigma^d)$ is the average number of obstacles for the system size and density considered. These $N$ obstacles are then placed uniformly at random within a hypersphere shell of inner radius $\sigma=1$ and outer radius $r_\mathrm{max} > \sigma$.
Because $N$ is chosen in accordance to the fluctuation of the Poisson random field in a finite volume, this construction guarantees that the probability of obtaining a cavity containing the origin, $\mathbb{C}$, exactly tracks the distribution of cavities at that same $\hat\varphi$ in an infinitely large system. The properties of this cavity can then be sampled using either static or dynamical algorithms.

\subsection{Static Sampling}
For a purely geometric sampling of the cavity properties, a Delaunay triangulation (into $d$-simplicial cells) of the obstacles within that cavity is built using CGAL's $d$D Triangulation library~\cite{cgal:hdj-t-19a}.
The cavity is then constructed by a graph search with cells as vertices and facets as edges. Starting from the cell that contains the origin, an edge (facet) is connected if the circumcenter of two cells are in same side, or the circumcenter are on opposite sides and the  circumradius of the facet is greater than $\sigma$. All visited cells are added to the cavity. The cavity is valid if the displacement of any sites in the cavity to the origin is less than $r_\mathrm{max}-\sigma$. Care must be taken choosing $r_\mathrm{max}$, such that this condition is met.
Like in the Leath algorithm for lattice percolation~\cite{leath1976clustersize}, cavities are evenly sampled in a site base, that is, the probability of generating a cavity of volume $V_\mathrm{cavity}$ is proportional to $V_\mathrm{cavity} P(V_\mathrm{cavity})$, where $P(V_\mathrm{cavity})$ is the probability of having a cavity of volume $V_\mathrm{cavity}$ in the thermodynamic limit.

Sastry \textit{et al.} proved that the visited cells constructed this way contain and only contain the void space that belongs to the same cavity~\cite{sastry1997statistical}. They also introduced an exact algorithm to determine the cavity volume through a recursive division of $d$-simplices. Because the exact decomposition of a cavity into simple primitives is quite involved in general dimension, we consider instead a random sampling algorithm.
The basic idea is to generate uniformly distributed random points (samples) within the cavity and to use these samples to approximate the cavity volume and other physical quantities. The high level description of the principal algorithm is as follows:

\begin{algorithm}[H]
\caption{Sampling a cavity}
\begin{algorithmic}
\For{$C_i$ in visited cells}
  \State $V_i \leftarrow$ \Call{SimplexVolume}{$C_i$}
  \State Increment $V_\mathrm{cells}$
\EndFor
\For{$j=1$ to $N_\mathrm{samples}$}
  \State Randomly choose a simplex $C_k$ in $\{C_i\}$ with probability $V_k/V_\mathrm{cells}$
  \State Place a random sample $S \leftarrow$ \Call{SampleSimplex}{$C_k$}
  \If{$S$ in the void space}
    \State Add $S$ to the void sample list and increment $N_\mathrm{voids}$
  \EndIf
\EndFor
\end{algorithmic}
\end{algorithm}

Note that the volume of a $d$-simplex defined by the vertices $\{p_0, p_1, ..., p_d\}$ is
\begin{equation}
V_\mathrm{simplex} = \left| \frac{1}{d!} \det(p_1-p_0, p_2-p_0, ..., p_d-p_0) \right|.
\end{equation}

Obtaining uniform samples in a $d$-simplex is equivalent to generating $d+1$ random spacing with unit sum~\cite[p.~568]{devroye1986nonuniform}. To generate $d+1$ random spacings ,$x_0, ..., x_d$, one first generates $d$ independent and uniformly distributed random variables $y_1, ..., y_d$ in $[0, 1)$ and sort them in place, in addition to $y_0 = 0$ and $y_{d+1} = 1$, then $x_i = y_{i+1} - y_i$, and the random sample $S = \sum_{i=0}^d x_i p_i$.

Determining whether $S$ is in the void space requires a nearest-neighbor query of the obstacles. Although the nearest obstacle of $S$ is most likely to be one of the vertices of $C_i$, outliers are possible. To accelerate the computation, one may pre-compute the point-to-simplex distances of these possible obstacles other than the simplex vertices, and store those with distance less than $\sigma$ as candidate nearest neighbors.

As the obstacle density increases, the fraction and size of the voids become increasingly small, which makes this sampling approach inefficient. We then instead find the vertices of the cavity, build the triangulation over these vertices, and then run the cavity sampling algorithm in the new triangulation. Note that a simplex generated this way may lie completely in occupied space, or even contain the voids of other cavities. The later case should be rare and in fact is not observed in practice. One should nonetheless test for this case and drop occupied or invalid simplices from thus sampling. With this simple optimization, the fraction of void samples ($N_\mathrm{voids}/N_\mathrm{samples}$) typically varies from a half to nearly one.

From the $N_\mathrm{voids}$ samples out of $N_\mathrm{samples}$ within the cavity, we approximate the cavity volume
\begin{equation}
V_\mathrm{cavity} = V_\mathrm{cells} \frac{N_\mathrm{voids}}{N_\mathrm{samples}}.
\end{equation}
where $V_\mathrm{cells}$ is the total volume of visited cells. From the set of samples within the void space, $\{S_i\}$, we can also approximate the infinite-time mean squared displacement of a tracer in this cavity as
\begin{equation}
\Delta_\mathrm{cavity} = \langle (S_i - S_j)^2 \rangle = 2 (\langle S_i^2 \rangle - \langle S_i \rangle^2 ).
\end{equation}
The self van Hove function, $G_\mathrm{s}(r, t)$, which is defined as the probability of finding a tracer at displacement $r$ at time $t$, can be computed in the $t \rightarrow \infty$ limit, when every site is equally probable, for a single cavity as
\begin{equation}
G_\mathrm{s,cavity}(r) = G_\mathrm{s,cavity}(r, t \rightarrow \infty) \sim \sum_{i \neq j} \delta(|S_i - S_j|-r)
\end{equation}
and is normalized as $\int_0^\infty G_\mathrm{s}(r) \dd r = 1$. Note that the summation is over sites $i \neq j$, in order to eliminate the artificial peak at $r = 0$ due to the discretization scheme.
Finally, the expected $V$, $\Delta$ and $G_\mathrm{s}(r)$ are the arithmetic mean over all randomly generated cavities.

\subsection{Dynamical Sampling} \label{sec:rlgdynamics}

For the tracer dynamics, we implemented a high-dimensional generalization of the simulation scheme of H{\"o}fling \emph{et al.}~\cite{hofling2006localization,hofling2008critical}.

\begin{figure*}[ht]
\includegraphics[width=\textwidth]{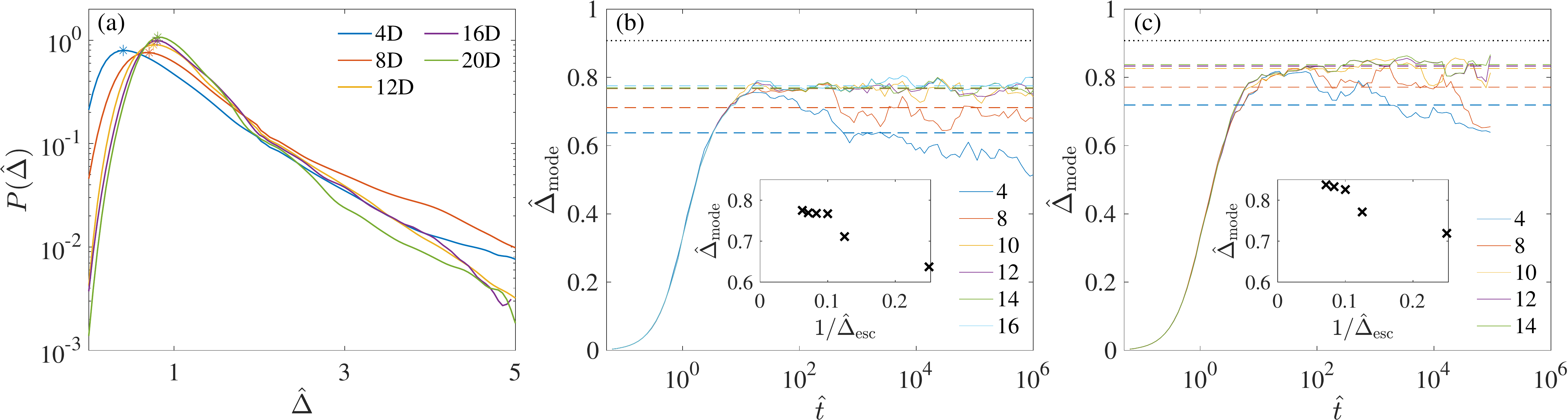}
\caption{(a) Dynamical cage size distribution in $\hat\varphi = 2.7$ at $\hat{t} = 2^{16}$ in various dimensions. The modal cage size $\hat\Delta_\mathrm{mode}$ is denoted by asterisks. (b, c) $\hat\Delta_\mathrm{mode}(t)$ in $\hat\varphi=2.7$ and (b) $d=16$ and (c) $d=20$ obtained for systems with different shell thickness $\hat\Delta_\mathrm{max}$. Colored dashed lines denote the plateau value of $\hat\Delta_\mathrm{mode}$ and also plotted in insets. The black dotted line denotes the mean-field prediction.}
\label{fig:sitypical}
\end{figure*}

For the cage escape analysis, obstacles are first generated according to the cavity reconstruction scheme, which allows the vicinity of  $\hat\varphi_\mathrm{d}$ in dimensions as high as $d=20$ to be reached.
A tracer is then placed at the origin and ballistic dynamics is run. The simulation terminates when $t_\mathrm{max}$ is reached or when the tracer escapes the shell, i.e., $r(t) > r_\mathrm{max} - \sigma$, whichever comes first. The maximal valid tracer square displacement $\hat\Delta_\mathrm{max} = d \times (r_\mathrm{max} - \sigma)^2$ defines the simulation shell thickness. For a given $t$, multiple time intervals are sampled and averaged to obtain the dynamical cage size $\Delta(t)$ for a specific realization of disorder. 
The mode cage size at time $t$ is defined as the maximum likelihood value in the distribution of time dependent cage sizes, $\hat\Delta_\mathrm{mode}(t) = \arg\max P(\hat\Delta(t))$ (Fig.~\ref{fig:sitypical} (a)). 
The escape event time, $t_\mathrm{esc}$ at $\Delta_\mathrm{esc}$ is calculated as the first-passage time of the tracer square displacement from the origin being $\Delta_\mathrm{esc}$.

By construction, the cavity reconstruction scheme exhibits no finite-size correction as long as the cage is closed. The obstacles are indeed then generated according to the Poisson distribution, as in an infinite-size system.
For imperfectly closed cages, however, finite-size corrections arise because the tracer escapes that cage at different times, depending on the shell thickness. While the MSD is sensitive to the rare samples that exhibit large displacement, we find that finite-size corrections to $\hat\Delta_\mathrm{mode}$ only become larger than statistical noise for fairly small system sizes, e.g., $\hat\Delta_\mathrm{max} \le 8$ in $\hat\varphi=2.7$ and $d=16$. A significant fraction of tracers can then escape on a time comparable to reaching the plateau height, and hence $\hat\Delta_\mathrm{mode}(t)$ shrinks with time. In all other cases, the mode reaches a plateau that persists for multiple time decades. We thus extract the plateau value of $\hat\Delta_\mathrm{mode}$ by taking the average of $\hat\Delta_\mathrm{mode}(\hat{t})$ from $\hat{t}=50$ to $10^5$.
The magnitude of the finite-size effect is comparable with the statistical noise of extracting the modal cage size from different realization of sample cages.
It is worth note that, because in the highest approachable dimension, $d=20$, a smaller simulation shell size, $\hat\Delta_\mathrm{max}=14$, is used than in $d\leq16$, the difference between the plateau heights, which scales like $1/d$, then becomes statistically indistinguishable~(Fig.~4(a) in the main text).

To assure better accuracy, we randomly choose $4000$ samples out of the total, evaluating the modal cage size and repeat multiple times. This bootstrap sampling gives the expected $\hat\Delta_\mathrm{mode}$ and the confidence interval of the approximation shown in Fig.~4(b) in the main text. Results are then fitted with linear form with zero intercept. We chose this intercept in order to obtain clearer results for Fig. 4(c). If the intercept is fitted as well, its value deviates at most $\pm 5\%$ from the origin, which is well within the accuracy of the fitted data.

\bibliography{abbrev}

\end{document}